\documentclass[runningheads]{llncs}
\usepackage[T1]{fontenc}
\usepackage[disable]{todonotes}
\usepackage{subcaption}

\usepackage{orcidlink}

\usepackage{color}

\usepackage{cite}
\usepackage{amsmath,amssymb,amsfonts}
\usepackage{algorithmic}
\usepackage{graphicx}
\usepackage{textcomp}
\usepackage{xcolor}
\def\BibTeX{{\rm B\kern-.05em{\sc i\kern-.025em b}\kern-.08em
    T\kern-.1667em\lower.7ex\hbox{E}\kern-.125emX}}

\usepackage[most]{tcolorbox}
\usepackage{listings}
\usepackage{ebproof}
\usepackage{amssymb}
\usepackage{mathtools}
\usepackage{multirow}

\usepackage{esvect}
\usepackage{breqn}
\usepackage{breakcites}
\usepackage{stmaryrd}

\usepackage{amsmath}

\usepackage{caption}

\usepackage{xspace}

\usepackage{enumitem}

\usepackage{adjustbox}

\lstdefinelanguage[RISCV]{Assembler}
{
  alsoletter={.}, 
  alsodigit={0x}, 
  morekeywords=[1]{ 
    lb, lh, lw, lbu, lhu, mv,
    sb, sh, sw, mul, addi, li,
    sll, slli, srl, srli, sra, srai,
    add, addi, sub, lui, auipc,
    xor, xori, or, ori, and, andi,
    slt, slti, sltu, sltiu,
    beq, bne, blt, bge, bltu, bgeu,
    j, jr, jal, jalr, ret,
    scall, break, nop
  },
  morekeywords=[2]{ 
    .align, .ascii, .asciiz, .byte, .data, .double, .extern,
    .float, .globl, .half, .kdata, .ktext, .set, .space, .text, .word
  },
  morekeywords=[3]{ 
    zero, ra, sp, gp, tp, s0, fp,
    t0, t1, t2, t3, t4, t5, t6,
    s1, s2, s3, s4, s5, s6, s7, s8, s9, s10, s11,
    a0, a1, a2, a3, a4, a5, a6, a7,
    ft0, ft1, ft2, ft3, ft4, ft5, ft6, ft7,
    fs0, fs1, fs2, fs3, fs4, fs5, fs6, fs7, fs8, fs9, fs10, fs11,
    fa0, fa1, fa2, fa3, fa4, fa5, fa6, fa7
  },
  morecomment=[l]{;},   
  morecomment=[l]{\#},  
  morestring=[b]",      
  morestring=[b]'       
}

\lstdefinelanguage{SML}
{
  morekeywords= {
    Definition, Theorem, Proof, QED, End, EQUAL, GREATER, LESS, NONE, SOME, abstraction, abstype, and, andalso, array, as, before, bool, case, char, datatype, do, else, end, eqtype, exception, exn, false, fn, fun, functor, handle, if, in, include, infix, infixr, int, let, list, local, nil, nonfix, not, o, of, op, open, option, orelse, overload, print, raise, real, rec, ref, sharing, sig, signature, string, struct, structure, substring, then, true, type, unit, val, vector, where, while, with, withtype, word
  },
  morestring=[b]",
  morecomment=[s]{(*}{*)},
  keywordstyle=\bfseries
}

\newcommand{\smartparagraph}[1]{\smallskip \noindent \textbf{#1.}}

\newcommand \keyfont {\mathsf}

\newcommand{\registername}[1]{\texttt{#1}}

\newcommand{\bstmtassign}[2]{{#1}\keyfont{\ :=\ }{#2}}
\newcommand{\bstmtassert}[1]{\keyfont{assert\ }{#1}\keyfont{\ }}
\newcommand{\bstmtcjmp}[3]{\keyfont{if\ }{#1}\keyfont{\ goto\ }{#2}\keyfont{\ else\ }{#3}}

\newcommand{\intprdom}[1]{\mathit{Dom}(#1)}
\newcommand{\freshsym}[1]{\mathit{fs}(#1)}

\newcommand{\Bexps}{\mathit{Exp}}
\newcommand{\Bexpsval}{e}

\newcommand{\Beld}[2]{\keyfont{ld(}#1\keyfont{,}\ #2\keyfont{)}}
\newcommand{\Best}[3]{\keyfont{st(}#1\keyfont{,}\ #2\keyfont{,}\ #3\keyfont{)}}

\newcommand{\Bvars}{\mathit{BVars}}
\newcommand{\Bvarsval}{x}
\newcommand{\BV}{\mathit{BV}}

\newcommand{\BVval}{v}

\newcommand{\Benv}{\delta}

\newcommand\pfun{\mathrel{\ooalign{\hfil$\mapstochar\mkern5mu$\hfil\cr$\to$\cr}}}
\newcommand{\BenvupdMap}[2]{#1 \mapsto #2}

\newcommand{\Benvupd}[3]{#1[\BenvupdMap{#2}{#3}]}

\newcommand{\expeval}[2]{{\llbracket #1 \rrbracket_{#2}}}

\newcommand{\state}{s}
\newcommand{\states}{S}
\newcommand{\stateerr}{\bot_{\state}}
\newcommand{\lbl}{l}
\newcommand{\lbls}{L}

\newcommand{\pc}[1]{\mathit{pc}(#1)}
\newcommand{\env}[1]{\mathit{store}(#1)}
\newcommand{\transition}[2]{{#1} \rightarrow {#2}}

\newcommand{\transitionf}[3]{{#1} \rightsquigarrow_{#2} {#3}}
\newcommand{\transitionsf}[4]{{#1} \rightarrow^{#2}_{#3} {#4}}

\newcommand{\prog}{p}

\newcommand{\symb}[1]{\overline{#1}}

\newcommand{\symbbenvinit}[0]{\symb{\Benv}_{\mathit{id}}}

\newcommand{\Bsyms}{\mathit{BSyms}}
\newcommand{\Bsymsval}{\alpha}

\newcommand{\bspcond}{\phi}

\newcommand{\bspath}[1]{\mathit{path}(#1)}

\newcommand{\bsresjmptgt}[3]{{t}_{{#3}}}

\newcommand{\Bsyminterp}{H}

\newcommand{\sexpeval}[2]{{\llparenthesis #1 \rrparenthesis_{#2}}}

\newcommand{\symmatch}[3]{#1 \mathrel{\stackrel{\makebox[0pt]{\mbox{\normalfont\tiny #2}}}{\triangleright}} #3}
\newcommand{\symmatchext}[3]{#1 \mathrel{\stackrel{\makebox[0pt]{\mbox{\normalfont\tiny #2}}}{\blacktriangleright}} #3}

\newcommand{\minimInterpr}[2]{\downarrow_{#1}({#2})}

\newcommand{\EsymbexecH}{\hookrightarrow}
\newcommand{\Esymbexec}[2]{{#1}\EsymbexecH{#2}}
\newcommand{\EsymbexecF}[1]{\mathit{symsem}({#1})}

\newcommand{\HLsymbexecLeft}[2]{#2 \rightsquigarrow_{#1}}
\newcommand{\HLsymbexec}[3]{\HLsymbexecLeft{#1}{#2} #3}
\newcommand{\HLsymbexecStruct}{\Pi}

\newcommand{\SyStateSymbSubst}[3]{#1[#3 / #2]}

\newcommand{\BsyminterpPHolds}[2]{#1(#2)}
\newcommand{\SymbolsOf}[1]{\Lambda(#1)}

\newcommand{\opstep}{symbstep}
\newcommand{\opcase}{case}
\newcommand{\opinf}{infeasible}
\newcommand{\opfreesymbren}{freesymb\_rename}
\newcommand{\opren}{rename}
\newcommand{\opfreshif}{simplify}
\newcommand{\opsubst}{subst}
\newcommand{\opconseq}{consequence}
\newcommand{\optransf}{transfer}
\newcommand{\opseq}{sequence}

\newcommand{\opref}[1]{\textsc{#1}}
\newcommand{\optabref}[1]{\textsc{#1}}

\def\@fnsymbol#1{\ensuremath{\ifcase#1\or *\or \dagger\or \ddagger\or
   \mathsection\or \mathparagraph\or \|\or **\or \dagger\dagger
   \or \ddagger\ddagger \else\@ctrerr\fi}}

\newcommand{\ssymbol}[1]{^{\@fnsymbol{#1}}}

\newcommand{\toolchain}{HolBA-SE\xspace}

\newcommand{\revision}[5]{#5}

\hypersetup{hidelinks}

\begin{document}

\title{Forward Symbolic Execution for Trustworthy Automation of Binary Code Verification\thanks{This preprint has not undergone peer review (when applicable) or any post-submission improvements or corrections. The Version of Record of this contribution is published in ``Proceedings of the 27th International Conference on Verification, Model Checking, and Abstract Interpretation – VMCAI 2026'', and is available online at \url{https://doi.org/10.1007/978-3-032-15700-3_8}.}}
\titlerunning{Forward Symbolic Execution for Binary Code Verification}
%
\author{
Andreas Lindner\inst{1}\orcidlink{0000-0001-5311-1781}
\and
Karl Palmskog\inst{2}\orcidlink{0000-0003-0228-1240}
\and
Scott Constable\inst{3}\orcidlink{0000-0002-6793-0451}
\and
Mads Dam\inst{2}\orcidlink{0000-0001-5432-6442}
\and
\\Roberto Guanciale\inst{2}\orcidlink{0000-0002-8069-6495}
\and
Hamed Nemati\inst{2}\orcidlink{0000-0001-9251-3679}
}
\authorrunning{Lindner et al.}
%
\institute{
Uppsala University, Sweden\\
\email{andreas.lindner@angstrom.uu.se}
\and
KTH Royal Institute of Technology, Sweden\\ \email{\{palmskog,mfd,robertog,hnemati\}@kth.se}
\and
Intel Labs, United States\\
\email{scott.d.constable@intel.com}
}
\maketitle              
\begin{abstract}
Control flow in unstructured programs can be complex and dynamic, which makes static analysis difficult.
Yet, automated reasoning about unstructured control flow is important when certifying properties of binary (machine) code in trustworthy systems, e.g., cryptographic routines. We present a theory of forward symbolic execution for unstructured programs suitable for use in theorem provers that enables automated verification of both functional and non-functional program properties. The theory's foundation is a set of inference rules where each member corresponds to an operation in a symbolic execution engine. The rules are designed to give control over the tradeoff between the preservation of precision and introduction of overapproximation.
We instantiate our theory for BIR, a previously proposed intermediate language for binary analysis. We demonstrate how symbolic executors can be constructed for BIR with common optimizations such as pruning of infeasible symbolic states.
We implemented our theory in the HOL4 theorem prover using the HolBA binary analysis library, obtaining machine-checked proofs of soundness of symbolic execution for BIR. 
We practically evaluated two applications of our theory: verification of functional properties of RISC-V binaries and verification of execution time bounds of programs running on the ARM Cortex-M0 processor. The evaluation shows that such verification can be automated with moderate overhead on medium-sized programs.

\end{abstract}
%
%
%


\section{Introduction}
\todo[inline]{
(limit: 20 pages)

intro: now 2.7p, aim 2p
background: now 3p, aim 3p
symbsem: now 1p, aim 1p
symbrules (main contribution): now 6p (example 1.5p), aim 6p (maybe 1 page for the example)
implementation/evaluation: now 2.5p, aim 2p
application riscv: now 3.5p, aim 2.5p
application wcet: now 1.7p, aim 1.5p
related work+conclusion: now 2.3p, aim 1.5p (or less)

----

----

Section         Status      TODO
Intro           +           Review , shorten if possible {Hamed}
Background      +           Fix Figure, Fix reference to example, Review {Hamed}
Symb Semantics  +           Ok
Symb Execution  ~           4.1 4.2 To review {Roberto}, 4.3 to rewrite (to 1 page) {
                                Roberto initial Feedback Today,
                                Andreas new version by 4th at 07:00}
Implementation  -           TODO {Karl structure by 4th, Andreas rewrite/review by Friday}
Risc-V          +           Review {Hamed}
WCET            ~           7.1 adapt to example, 7.2 refine text {Andreas}

Deadline Friday at 13:00
Meet Tuesday    at 09:00

Rel Work        ~           To revise (shrink/tone down+cite crypto se https://arxiv.org/pdf/2505.14348)
Conclusions     ~           To revise
}

Unstructured programming permits ``jumps of the control to strange places'' via labels and goto instructions, resulting in control flow with the appearance of a ``can of spaghetti''~\cite{Hamming1997}. 
Structured programming~\cite{Dahl1972}, where instructions or statements have a single exit point, is now the norm for code written by software developers---but code with unstructured control flow lives on in high assurance systems, e.g., as output of compilers and as hand-optimized assembly.

Static analysis of unstructured code, e.g., of RISC-V binaries~\cite{riscv}, is difficult due to its complex, dynamic control flow. For instance, jump targets may be computed at runtime from register and memory values (\emph{indirect jumps}).
%
%
Nevertheless, trustworthy \emph{automated} reasoning about binary code is important for certifying the correctness of high-assurance systems. In particular, binary verification may be used for \emph{translation validation} when using untrusted compilers~\cite{Sewell2013} and for directly ensuring the correctness of low-level cryptographic code~\cite{Mazzucato2025} and security mechanisms such as software fault isolation~\cite{zhao2011armor}.




We believe \emph{forward symbolic execution} is well-suited for automating binary verification. Such an approach
explores program paths with symbolic input, naturally captures cause-effect in control and data flow, tracks how data drives behavior without source-level types, and integrates well with control-flow reconstruction and memory-alias analysis. 
Based on these insights, we present a general theory of forward symbolic execution for unstructured programs suitable for use in interactive theorem provers. 
The basis of the theory is a set of core inference rules, each corresponding to operations in an abstract symbolic execution engine, designed to control the tradeoff between precision and overapproximation, which is important when analyzing real-world binary code (Sec.~\ref{nfm2023:sec:symsem} and~\ref{sec:symexec}).
The theory can be instantiated for unstructured languages by relating their concrete and symbolic semantics. We instantiate the theory for BIR, a previously proposed intermediate language for binary analysis~\cite{Lindner2019}, and construct sound symbolic executors with optimizations such as pruning of infeasible states.
We implemented~\todo{MFD: Don't like the transition to past tense here} the theory in the HOL4 theorem prover~\cite{hol4} and instantiated it using\todo{AL: 'and instantiated for' to separate the rule implementation from BIR instance in this sentence? also theory vs executor implementation} the HolBA binary analysis library~\cite{Lindner2019}, providing machine-checked proofs of soundness of symbolic execution for BIR (Sec.~\ref{nfm2023:sec:impl}).

Using our implementation, we extended HolBA with support for two novel applications: (1)~automated verification of functional properties of RISC-V binaries using contracts and (2)~automated verification of execution time bounds of binaries executing on the ARM Cortex-M0 processor. We call this extension \toolchain.
Our application of \toolchain for functional verification of {RISC-V} code is based on previous work which \emph{lifts} RISC-V instructions to BIR~\cite{Lindner2019} while preserving semantics as defined by L3~\cite{Fox2015}, and provides Hoare-style contracts for binary programs~\cite{Lundberg2020}. 
To this we add trustworthy automation of BIR contract proofs and for propagating such proofs to RISC-V level (Sec.~\ref{sec:riscv-application}). We applied \toolchain to RISC-V case studies from two high-assurance domains, namely, cryptography and operating system kernels, indicating that we can automate translation validation and verification of hand-written assembly outside the scope of verified compilers.
In our application of \toolchain for verification of execution time bounds (Sec.~\ref{nfm2023:sec:impl:exectime}), we rely on the fact that in relatively simple processors such as the ARM Cortex-M0, instructions take constant time to execute. This means that we can explicitly represent execution time as a symbolic state variable, which we use to perform a collection of case studies on ARM Cortex-M0 binaries where we verify their execution time bounds. Our evaluation compares our results against an established industrial tool, aiT, showcasing the potential of our approach for trustworthy worst-case execution time (WCET) analysis.

In summary, we make the following contributions:
\begin{enumerate}
\item We propose a general theory of symbolic execution for unstructured programs designed and adapted for use in interactive theorem provers.
\item We provide an implementation of our theory in the HOL4, along with an instantiation for the BIR language and machine-checked soundness proofs.
\item We design symbolic executors that automatically apply our theory's inference rules for symbolic execution of BIR programs generated from RISC-V and Cortex-M0 machine code, building HOL4 progress structure theorems.
\item Using progress structures, we implement automation for functional verification of RISC-V programs and WCET verification for Cortex-M0 programs on top of the HolBA library, experimentally evaluated in several case studies.
\end{enumerate}

Source code for \toolchain and our verification case studies is available under an open source license~\cite{HolBA-SE}, and has been integrated into the HolBA library~\cite{HolBA}.

\section{Background}
\label{sec:background}

We use BIR~\cite{Lindner2019} as a vehicle for explaining our theory and implementation. BIR (Binary Intermediate Representation) is an intermediate language designed for binary analysis. It abstracts away ISA (Instruction Set Architecture) details while preserving the semantics of binaries, enabling architecture-independent reasoning about machine code. Beyond serving as a convenient notation in this paper, BIR provides a general framework for implementing trustworthy binary analyses, thanks to its formal semantics and implementation in the HOL4 theorem prover.
The application of symbolic execution to binary code presents unique challenges. To illustrate these, we use a modular exponentiation (ModExp) routine expressed in BIR, whose control flow graph is presented in Fig.~\ref{fig:modexp_imp_cfg}. 
The pseudocode of the corresponding modular exponentiation algorithm is shown in Fig.~\ref{fig:modexp_imp_code}. For simplicity, the program operates only on 32-bit words instead of relying on a bignum library.
The program highlights key challenges---unstructured control flow, path explosion, expression growth, and memory aliasing---that motivate our symbolic execution rules and optimizations.

\subsection{The BIR language}
\label{sec:background:bir}

\newcommand{\fstlangcol}[1]{#1}


A BIR program consists of uniquely labeled blocks, with each block containing a sequence of statements.
In the running example, the blocks consist of a single statement and are labeled by $1,\dots,14$.
Labels $\lbl \in \lbls_\prog$ correspond to specific locations in the program $\prog$ and are commonly used as the target for jump instructions. 
BIR statements include
assignment of BIR expressions to variables (e.g., \verb|SP := SP - 8|), 
unconditional and conditional jumps (e.g., \verb|goto R3| \todo{RG: decide to use goto or jmp} and \verb|if R3=32 goto 6|), 
a \verb|halt| statement to indicate execution termination, and 
\verb|assert| to evaluate a boolean expression and terminate execution if the assertion fails. 
Expressions in BIR include constants $\BVval \in \BV$, variables $\Bvarsval \in \Bvars$, conditionals, arithmetic operations, and memory operations such as \textit{load} ($\verb|ld|$) and \textit{store} ($\verb|st|$).

\label{nfm2023:app:concsem}

Fundamentally, the content of the following sections is not specific to BIR, but requires the underlying concrete model 
and its semantics to satisfy some properties, which indeed are satisfied by BIR.
The concrete execution is modeled by a deterministic and total transition relation $\transition{}{}$. In BIR, states $\perp$ and $\top$ (reached by failed assertions and halt, respectively) represent failing and successful terminating states that can only transition to themselves;
every concrete state $\state \in \states$ consists of a 
program counter and a store, extracted by $\pc{\state}$ and $\env{\state}$, respectively;
Stores $\Benv$ partially map variables $\Bvarsval \in \Bvars$ to concrete values $\BVval \in \BV$, and
instructions are addressed using labels ranged over by $\lbl$.

Because code is unstructured, we use sets of labels $\lbls \subseteq \lbls_\prog$ (not necessarily consecutive)
to identify fragments of binary code.
We define $\transitionf{\state}{\lbls}{\state'} = \{ (\state, \state') \mid \transition{\state}{\state'} \land \pc{\state} \in \lbls \}$ as the restriction of the
transition relation\revision{AL}{R2.P4}{}{to source states that}{, where source states must} belong to a fragment $\lbls$, and $\transitionsf{}{n}{\lbls}{}$ for the
application of $n$ transitions.
Note that these relations are partial and deterministic, and if
$\state_0 \transitionsf{}{n}{\lbls}{} \state_{n}$
then each intermediate state has program counter in $\lbls$: if $i < n$ and $\transitionsf{\state_0}{i}{\lbls}{\state_i}$ then  $\pc{\state_i} \in \lbls$.

\subsection{Running example} 
\label{sec:background:example}

\todo{RG: Commented story on secondary goals (i.e., CFI), these should be moved in intro as motivation}

\begin{figure}[t]
\centering
\begin{minipage}[t]{0.35\textwidth}
  \vspace{0pt} 
  \centering
\begin{lstlisting}[mathescape, basicstyle=\small\ttfamily, language=C, frame=single]
f(e, b, m):
 [prologue]
 r := 1
 for i in 0 .. 31 {
   if e[i]
      r := (r*b)%m
   b := (b*b)%m
 }
 return r
 [epilogue]
\end{lstlisting}
  \captionof{figure}{ModExp pseudocode.}
  \label{fig:modexp_imp_code}
\end{minipage}
\begin{minipage}[t]{0.05\textwidth}
  \vspace{0pt}
\  
\end{minipage}
\begin{minipage}[t]{0.56\textwidth}
  \vspace{0pt} 
  \centering
  \includegraphics[width=\linewidth]{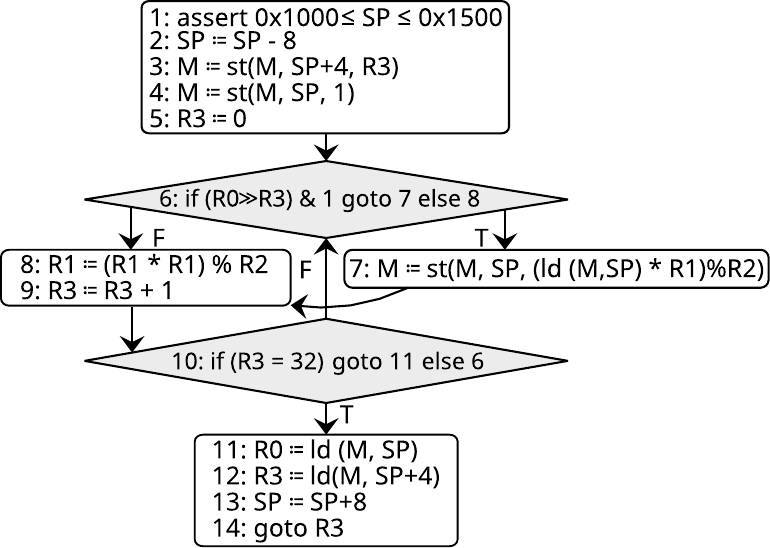}
  \captionof{figure}{BIR CFG of ModExp.}
  \label{fig:modexp_imp_cfg}
\end{minipage}
\end{figure}

\noindent The BIR program in Fig.~\ref{fig:modexp_imp_cfg} implements modular exponentiation
using the standard right-to-left method, computing $b^e \bmod m$. Parameters $e$, $b$, and
$m$ are passed in BIR variables \texttt{R0}-\texttt{R2}, and \texttt{R3} stores the
return address. 
The first instruction (line 1) asserts that the stack pointer (\registername{SP}) is restricted to the designated stack region of memory, spanning addresses 0x1000 to 0x1500. This assumption becomes the initial path condition during symbolic execution.
Lines 2–3 form the function prologue: the function allocates two words to the stack and then pushes the return address (passed in through the R3 register ) onto the stack, whose top is pointed to by the stack pointer \registername{SP}. In BIR, memory is modeled as a single map variable $M$, which is accessed and modified by the operations \verb|ld| (load) and \verb|st| (store). The local variable $r$ is stored at address \texttt{SP - 8}, 
and line 4 sets its initial value to 1.
The loop counter $i$ is tracked in register \texttt{R3}, which is initialized to 0 at line 5. Line 6 is a conditional jump that evaluates the \texttt{if} condition and determines the subsequent control flow. Lines 7 and 8 implement the multiplication-and-modulo operations of the pseudocode: line 7 updates the memory location, where $r$ is allocated on the stack, while line 8 updates the base value $b$ directly in a register.
Lines 9 and 10 implement the loop’s iteration logic, incrementing the counter and branching back to line 6 until 32 iterations have been completed, after which execution proceeds to line 11. At this point, the value of $r$ is reloaded from the stack into the return register \texttt{R0}. Finally, lines 12–14 constitute the epilogue: the function restores the original return address, resets the stack pointer, and executes an indirect jump via \texttt{R3} to return control to the caller.

\subsection{Challenges and approach} 
\label{sec:background:approach}
This example highlights several core difficulties of symbolic execution for binary code:
\textbf{C1 Unstructured control flow}: Indirect jumps complicate reconstruction of feasible control-flow targets, since jump addresses may depend on runtime values in registers or memory. For instance, instead of branching directly at line 6, the program could implement the conditional jump via a lookup table stored in memory, where the target address is computed from the branch condition, e.g., \verb|goto ld(M, 0x100+(((R0>>R3)&1)*4))|. In this case, the jump destinations (e.g., lines 7 and 8) are retrieved from memory addresses 0x104 and 0x100, respectively. Such table-based indirection introduces an additional layer of complexity, requiring the analysis to identify all feasible addresses and corresponding memory contents to recover the possible control flow.
\textbf{C2 Path explosion}: The conditional branch inside the loop leads to an exponential number of feasible execution paths, in this case $2^{32}$).
\textbf{C3 Expression growth}: Operations that repeatedly manipulate memory and registers, such as the modular multiplication in line 7, duplicate symbolic expressions, and cause exponential expression growth.
\textbf{C4 Memory aliasing}: Since binary programs operate on flat memory, reasoning about disjointness of memory locations becomes necessary, for example, to show that internal loop operations do not overwrite the return address.
\textbf{C5 Repeated fragments}: Loops and repeated function calls execute multiple times the same code fragments, leading to redundant symbolic computations if treated naively.
\textbf{C6 Interval reasoning}: Many analyses require reasoning about bounded counters or index values (e.g., the loop variable), which is difficult to express without explicit support for symbolic intervals.
\textbf{C7 Architectural dependencies}: Symbolic execution at the binary level is tightly coupled to the ISA and its semantics.
Without an intermediate language like BIR, each symbolic executor must be tailored to a specific ISA, limiting reuse across architectures.

Our approach to \revision{AL}{R2.P5}{}{these challenges}{addressing these challenges simultaneously} is built on three key elements. \textbf{Forward exploration}: for unstructured code (C1), feasible targets of indirect jumps or table lookups can only be determined by symbolically executing the preceding instructions. 
\textbf{Compositional inference rules}: These rules support controlled overapproximation through the introduction of free symbols. The rules address C1 (path explosion) and C2 (expression blow-up) by merging equivalent states and bounding symbolic expression growth, while C4 (memory aliasing) is mitigated through simplification rules that identify equivalent or disjoint memory accesses. The rules also address C5 (repeated fragments) by allowing the instantiation of previously analyzed code fragments, and address C6 (interval reasoning) by manipulating path conditions and free symbols in a sound manner. 
\textbf{Integration with HolBA}: Binaries are lifted into BIR, an intermediate language with a formally verified transpiler in HOL4. \revision{AL}{R3.S2.3}{}{BIR}{HolBA} provides an ISA-independent substrate, directly addressing C7 by allowing the same symbolic executor to reason about RISC-V, ARM, or other architectures once their lifting to BIR is defined.

Our symbolic execution data structures and soundness definitions are designed to allow incremental proofs, facilitating implementation \revision{AL}{R1+2.interactive/automatic}{}{}{of proof automations} in interactive theorem provers.
\revision{AL}{R3.S2end}{}{As described in more detail in Sec.~\ref{nfm2023:sec:impl}, we implement parameterized execution blocks that compose into symbolic executors. Each executor instantiates inference rules within HOL4 to produce machine-checked theorems, making them reusable building blocks for verifying binary code properties.}{
As an instance, we describe our implementation of automatic symbolic executors and semi-automatic optimization procedures within HOL4 and showcase the verification of binary code properties.
}



\section{Symbolic Semantics}
\label{nfm2023:sec:symsem}
The inference rules for symbolic execution in Sec.~\ref{sec:symexec} are general, but require a symbolic semantics for individual instructions of the specific model.
In this section, we lay down notation for the symbolic semantics and we exemplify a few rules for the symbolic semantics of BIR, which captures the main traits of the semantics supported by our symbolic executor.

We assume symbols $\Bsymsval$ that range over the set $\Bsyms$ and a symbolic expression language
that supports at least first-order logic.
Symbolic expressions are denoted by $\symb{\BVval} \in \symb{\BV}$.
A symbolic store $\symb{\Benv}$ is a mapping of variables to symbolic expressions.
We use  $\symbbenvinit$ to refer to the symbolic store that maps each variable to a different symbol, i.e.,  $\symbbenvinit(\registername{V}) = \Bsymsval_{\registername{V}}$ for each $\registername{V}$.
We use $\bspcond$ for boolean symbolic expressions, which are used, for instance, as path conditions.
A symbolic state $\symb{\state}$ is a triple of a concrete program counter $\lbl$, a symbolic store $\symb{\Benv}$, and a path condition $\bspcond$.
We define the accessor functions $\pc{\symb{\state}} = \lbl$, $\env{\symb{\state}} = \symb{\Benv}$, $\bspath{\symb{\state}} = \bspcond$.

The symbolic and concrete semantics are connected using a partial interpretation $\Bsyminterp : \Bsyms \rightarrow \BV$, mapping symbols to concrete values. 
Interpretation is lifted to symbolic expressions and stores, i.e., 
the interpretation $\Bsyminterp(\symb{\BVval})$ produces a concrete value, and $\Bsyminterp(\symb{\Benv})$ produces a concrete store.
\todo{RG: How about overapproximation? AL: this is how we still have defined it now in HOL4, however Theorem 1 allows to use an overapproximating definition of it (currently we use an oracle constructed from z3 for generating these overapproximating concretizations, that's why it works)}
The resolution function $\bsresjmptgt{}{}{\bspcond}(\symb{\BVval}) = \{ \BVval \mid \exists \Bsyminterp.\ \Bsyminterp(\bspcond) \land \Bsyminterp(\symb{\BVval}) = \BVval \}$ computes the set of all concretizations of the symbolic expression $\symb{\BVval}$ under the path condition $\bspcond$.

\begin{definition}
  \label{nfm2023:def:symstatematch}
  The symbolic state $\symb{\state}$ \emph{matches} a concrete state $\state$ via the interpretation $\Bsyminterp$, written $\symmatch{\symb{\state}}{\Bsyminterp}{\state}$, if
  $\pc{\symb{\state}} = \pc{\state}$, $\Bsyminterp(\env{\symb{\state}}) = \env{\state}$, and $\Bsyminterp(\bspath{\symb{\state}})$.
\end{definition}
For sets of symbolic states $\HLsymbexecStruct$, we write $\symmatch{\HLsymbexecStruct}{\Bsyminterp}{\state}$ if there is an $\symb{\state} \in \HLsymbexecStruct$ such that  $\symmatch{\symb{\state}}{\Bsyminterp}{\state}$.

We exemplify the symbolic transition rule for conditional jumps in BIR:
\[
  \begin{prooftree}[separation=0.95em]
  	\hypo{\prog(\lbl) = \bstmtcjmp{\Bexpsval}{\Bexpsval_t}{\Bexpsval_f}}
	\hypo{
        \bspcond_t = \sexpeval{\Bexpsval}{\symb{\Benv}}
    \quad
        \bspcond_f = \neg{}\bspcond_t
    \quad
        x \in \{t,f\}
    }
	\hypo{\lbl_b \in \bsresjmptgt{}{\symb{\Benv}}{\bspcond \land \bspcond_x}(\sexpeval{\Bexpsval_x}{\symb{\Benv}})}
    \infer3[]{\prog \vdash \Esymbexec{(\lbl, \symb{\Benv}, \bspcond)}{
      (\lbl_b, \symb{\Benv}, \bspcond \land \bspcond_x)
     }}
  \end{prooftree}
\]
Note that the rule hypothesis requires knowing the current instruction.
Therefore, without loss of generality, we treat the program counter as a concrete value rather than a symbolic expression. In fact, the rules are defined to account for all possible target values of the program counter via the
resolution function. The rules use symbolic evaluation of the BIR expressions, which is written as $\sexpeval{\Bexpsval}{\symb{\Benv}}$ and substitutes in $\Bexpsval$ all variables with the corresponding symbolic expression from the store $\symb{\Benv}$ and results in a symbolic expression.
 
In the following, we use the total function
$\EsymbexecF{\symb{\state}} = \{\Esymbexec{\symb{\state}' \ \mid\ \symb{\state}}{\symb{\state}'}\}$ to represent all possible successors of an input state in the symbolic semantics.
Soundness means that the symbolic semantics overapproximates the concrete one, which we formulate as a simulation.
\begin{theorem}
\label{nfm2023:thm:sym_step_sound}
If $\symmatch{\symb{\state}}{\Bsyminterp}{\state}$ and
$\transition{\state}{\state'}$,
then exists $\symb{\state}' \in \EsymbexecF{\symb{\state}}$ such that $\symmatch{\symb{\state}'}{\Bsyminterp}{\state'}$.
\end{theorem}
Note that the same interpretation function $\Bsyminterp$ is used for matching both source and target states.
\section{Symbolic Execution}
\label{sec:symexec}
One could build the symbolic executor by iteratively applying $\EsymbexecF{\symb{\state}}$. However, this usually results in a large execution tree, which is infeasible to directly compute. Similarly, this naive approach neither allows for  overapproximations, which is
useful to speed up computation, nor to reuse previous results.
The main goal of the symbolic execution rules below is to formalize sound compositions, specializations, computation reuses, simplifications,  and approximations
that preserve soundness throughout the analysis. While we exemplify the rules only for the symbolic semantics of BIR, the
inference rule system has been proven correct for any symbolic and concrete semantics that satisfies Theorem~\ref{nfm2023:thm:sym_step_sound}.

\subsection{Progress structures}
The first building block is the formalization of the intermediate results of  symbolic execution.
Symbolic execution iteratively builds up and composes symbolic execution trees, but only keeps track of the roots and the leaves of the trees and disregards the internal nodes that are constructed with the symbolic semantics.
We call this object a progress structure and denote it as $\HLsymbexec{\lbls}{\symb{\state}}{\HLsymbexecStruct}$.
A progress structure captures that all concrete states matched by the source state $\symb{\state}$ execute through the label set $\lbls$ and eventually reach a concrete state that is matched by one of the target states in $\HLsymbexecStruct$.

In order to formally define the meaning of a progress structure, we first need to introduce some auxiliary notation.
The function $\SymbolsOf{\symb{\state}}$ returns the set of symbols that occur in $\symb{\state}$,
and we overload it for sets of states as the union of symbols in the states.
For a progress structure $\HLsymbexec{\lbls}{\symb{\state}}{\HLsymbexecStruct}$,
we call $bs(\HLsymbexec{\lbls}{\symb{\state}}{\HLsymbexecStruct}) = \SymbolsOf{\symb{\state}}$ its bound symbols 
and $\freshsym{\HLsymbexec{\lbls}{\symb{\state}}{\HLsymbexecStruct}} = \SymbolsOf{\HLsymbexecStruct} \setminus \SymbolsOf{\symb{\state}}$ its free symbols.
We use $\Bsyminterp' \supseteq \Bsyminterp$ to represent that the partial function $\Bsyminterp'$ extends $\Bsyminterp$, i.e. $\intprdom{\Bsyminterp'} \supseteq \intprdom{\Bsyminterp}$ and $\Bsyminterp'(\Bsymsval) = \Bsyminterp(\Bsymsval)$ for all $\Bsymsval \in \intprdom{\Bsyminterp}$.

The following two definitions support the formulation needed to overapproximate with fresh symbols in target states of progress structures.
\begin{definition}
\label{nfm2023:def:minim_interpr}
An interpretation $\Bsyminterp$ is minimal for $\symb{\state}$, written as $\minimInterpr{\symb{\state}}{\Bsyminterp}$, if $\intprdom{\Bsyminterp} = \SymbolsOf{\symb{\state}}$.
\end{definition}
This predicate is applied to source states to ensure that interpretations can be extended with arbitrary values for \emph{free} symbols not occurring in the initial state.
\begin{definition}
\label{nfm2023:def:extsymstatematch}
The symbolic state $\symb{\state}'$ loosely matches the concrete state $\state'$ via $\Bsyminterp$, written as $\symmatchext{\symb{\state}'}{\Bsyminterp}{\state'}$, if there is an interpretation $\Bsyminterp'\supseteq \Bsyminterp$ such that $\symmatch{\symb{\state}'}{\Bsyminterp'}{\state'}$.
\end{definition}
This predicate relaxes matching using interpretation extension to allow free symbols in target states $\symb{\state}'$ to take on arbitrary values in the matching of concrete states $\state'$.
%
%
%
%
We then define the meaning of a progress structure via its soundness.
\begin{definition}
\label{nfm2023:def:sym_exec_soundness}
A progress structure is sound, written $\vdash \HLsymbexec{\lbls}{\symb{\state}}{\HLsymbexecStruct}$, if
for every interpretation
$\minimInterpr{\symb{\state}}{\Bsyminterp}$
and state
$\symmatch{\symb{\state}}{\Bsyminterp}{\state}$, 
exists $\transitionsf{\state}{n}{\lbls}{\state'}$ where $n > 0$,
and exists $\symb{\state}' \in \HLsymbexecStruct$ such that
$\symmatchext{\symb{\state}'}{\Bsyminterp}{\state'}$.
\end{definition}
%
%
%
Compared with the soundness of the symbolic semantics (Theorem~\ref{nfm2023:thm:sym_step_sound}), the formulation of the soundness of the symbolic execution is different in two ways.

First, we assume minimality of the initial interpretation ($\minimInterpr{\symb{\state}}{\Bsyminterp}$) and allow loose matches for final states ($\symmatchext{\symb{\state}'}{\Bsyminterp}{\state'}$), which together enable the introduction of fresh symbols.
For example, suppose that the architecture has only registers \registername{R1} and \registername{R2}
and we symbolically execute line 8 of Fig.~\ref{fig:modexp_imp_cfg}. The symbolic semantics is $
\Esymbexec{(8, \symbbenvinit, T)}{(9, \Benvupd{\symbbenvinit}{\registername{R1}}{\Bsymsval_{\registername{R1}} * \Bsymsval_{\registername{R1}} \% \Bsymsval_{\registername{R2}}}, T)}$.
Progress structures allow \emph{forgetting} the exact value assigned to \registername{R1}, producing the sound progress structure
$\HLsymbexec{\{8\}}{(8, \symbbenvinit, T)}{\{(9, \Benvupd{\symbbenvinit}{\registername{R1}}{\Bsymsval'}, T)\}}$ (as long as $\Bsymsval' \not \in \SymbolsOf{\symbbenvinit}$).
Because we require any initial interpretation to be minimal (e.g., for a state $\{\registername{R1} \mapsto 2, \registername{R2} \mapsto 3\}$ the interpretation has to be $\{\alpha_{\registername{R1}} \mapsto 2, \alpha_{\registername{R2}} \mapsto 3\}$), this does not map the fresh symbol $\Bsymsval'$.
Instead, the loose interpretation matching allows to choose any extended interpretation (e.g. $\{\alpha_{\registername{R1}} \mapsto 2, \alpha_{\registername{R2}} \mapsto 3, \alpha' \mapsto 1\}$) that maps $\Bsymsval'$ to any value, including the value that the concrete execution produces.
If we had not assumed the initial interpretation to be minimal, the definition would be too strict and would not allow overapproximation with fresh symbols.
For example, it would require satisfying the property also for all interpretations that bind $\Bsymsval'$ to some other values, including the ones that do not match the final value of \registername{R1}.

Second, the concrete execution does not occur as an assumption but is a conjunct of the conclusion instead.
Therefore, a sound progress structure also guarantees the termination of the concrete execution:
the concrete source state must, in a finite number of transitions within $\lbls$, reach a concrete target state that is loosely matched by one of the target states.
We require $n > 0$ to be able to prove properties for looping code by ensuring progress when the source and target states point to the same label.

\todo[inline]{RG: The following sentence is obscure. AL: I attempted a fix, please check}
Unstructured code can be handled well with this soundness definition as arbitrary control flows are contained and abstracted as label sets, which represent code fragments.
\revision{AL}{R2.P9p2}{path sets would complicate loop merging, a question might be how much we limit the possibility of precision by ruling out the path detail, but then I think that we are not aiming in capturing the label paths anyways, we have the usual path condition as path representation so to say. I am not sure what the intuition of the reviewer is after her}{}{}
Note that indirect jumps are already handled by the symbolic semantics, which assumes knowledge about all feasible jump targets and captures those in the symbolic state set of Theorem~\ref{nfm2023:thm:sym_step_sound}.

A sound progress structure ensures that all matched concrete executions only encounter program labels in $\lbls$ before matching one of the targets $\symb{\state}' \in \HLsymbexecStruct$.
Therefore, if $\pc{\symb{\state}'} \notin \lbls$ then $\pc{\symb{\state}'}$ is only encountered once in the end.
For example, the progress structure from the previous paragraph guarantees that the program counter of the target state 9 is only encountered in the final execution state,
 because the relation $\transitionsf{\state}{n}{\lbls}{\state'}$ establishes that $\state'$ is reached and no transition is taken from a state pointing to $9 \notin \{8\}$.
This allows establishing properties when reaching a certain label of the program for the first time.
Because \revision{AL}{R3.S4.1}{}{the}{in Definition~\ref{nfm2023:def:sym_exec_soundness} the label set only appears in the} relation $\transitionsf{\state}{n}{\lbls}{\state'}$ \revision{AL}{R3.S4.1}{}{appears in the assumption of Definition~\ref{nfm2023:def:sym_exec_soundness} and extending its label set widens the relation, we may add}{, which is under an existential quantifier, extending label sets is permissible. We therefore may add} 9 to the label set of the previous progress structure and 
obtain a sound but weaker progress structure: $\HLsymbexec{\{8,9\}}{(8, \symbbenvinit, T)}{\{(9, \Benvupd{\symbbenvinit}{\registername{R1}}{\Bsymsval'}, T)\}}$.
In fact, since the program counter of the target state is now in the label set, according to Definition~\ref{nfm2023:def:sym_exec_soundness}, the concrete execution may go through loosely matching states several times, and thus potentially execute instructions 8 and 9 multiple times.


\begin{figure}[t]
  \centering
  \begin{minipage}[t]{0.46\textwidth}\vspace{0pt}
    \begin{prooftree}[separation=0.8em]
      \hypo{\EsymbexecF{\symb{\state}} = \HLsymbexecStruct}
      \hypo{\pc{\symb{\state}} \in \lbls}
      \infer2[\optabref{\opstep}]{\vdash \HLsymbexec{\lbls}{\symb{\state}}{\HLsymbexecStruct}}
    \end{prooftree}
  \end{minipage}\hfill
  \begin{minipage}[t]{0.54\textwidth}\vspace{0pt}
    \begin{prooftree}
      \hypo{\vdash \HLsymbexec{\lbls}{\symb{\state}}{\HLsymbexecStruct \cup \{(\lbl, \symb{\Benv}, \bspcond)\}}}
      \infer1[\optabref{\opcase}]{\vdash \HLsymbexec{\lbls}{\symb{\state}}{\HLsymbexecStruct \cup \{(\lbl, \symb{\Benv}, \bspcond \wedge \bspcond'), (\lbl, \symb{\Benv}, \bspcond \wedge \neg{\bspcond'})\}}}
    \end{prooftree}
  \end{minipage}

  \vspace{0.5em}

  \begin{minipage}[t]{0.48\textwidth}\vspace{0pt}
    \begin{prooftree}[separation=0.8em]
      \hypo{\vdash \HLsymbexec{\lbls}{\symb{\state}}{\HLsymbexecStruct \cup \{\symb{\state}'\}}}
      \hypo{\forall \Bsyminterp.\ \BsyminterpPHolds{\Bsyminterp}{\neg\bspath{\symb{\state}'}}}
      \infer2[\optabref{\opinf}]{\vdash \HLsymbexec{\lbls}{\symb{\state}}{\HLsymbexecStruct}}
    \end{prooftree}
  \end{minipage}\hfill
  \begin{minipage}[t]{0.42\textwidth}\vspace{0pt}
    \begin{prooftree}[separation=0.8em]
      \hypo{\vdash \HLsymbexec{\lbls}{\symb{\state}}{\HLsymbexecStruct}}
      \hypo{\Bsymsval' \notin \SymbolsOf{\symb{\state}} \cup \SymbolsOf{\HLsymbexecStruct}}
      \infer2[\optabref{\opren}]{\vdash \HLsymbexec{\lbls}{\SyStateSymbSubst{\symb{\state}}{\Bsymsval}{\Bsymsval'}}{\SyStateSymbSubst{\HLsymbexecStruct}{\Bsymsval}{\Bsymsval'}}}
    \end{prooftree}
  \end{minipage}

  \vspace{0.5em}

  \begin{prooftree}
    \hypo{\vdash \HLsymbexec{\lbls}{\symb{\state}}{\HLsymbexecStruct \cup \{\symb{\state}'\}}}
    \hypo{\Bsymsval \notin \SymbolsOf{\symb{\state}}}
    \hypo{\Bsymsval' \notin \SymbolsOf{\symb{\state}} \cup \SymbolsOf{\symb{\state}'}}
    \infer3[\optabref{\opfreesymbren}]{\vdash \HLsymbexec{\lbls}{\symb{\state}}{\HLsymbexecStruct \cup \{\SyStateSymbSubst{\symb{\state}'}{\Bsymsval}{\Bsymsval'}\}}}
  \end{prooftree}

  \vspace{0.5em}

  \begin{prooftree}
    \hypo{\vdash \HLsymbexec{\lbls}{\symb{\state}}{\HLsymbexecStruct}}
    \hypo{\Bsymsval \in \SymbolsOf{\symb{\state}}}
    \hypo{\SymbolsOf{\symb{\BVval}} \cap \freshsym{\HLsymbexec{\lbls}{\symb{\state}}{\HLsymbexecStruct}}  = \emptyset}
    \infer3[\optabref{\opsubst}]{\vdash \HLsymbexec{\lbls}{\SyStateSymbSubst{\symb{\state}}{\Bsymsval}{\symb{\BVval}}}{\SyStateSymbSubst{\HLsymbexecStruct}{\Bsymsval}{\symb{\BVval}}}}
  \end{prooftree}

  \vspace{-0.5em}
  \caption{Overview of the symbolic-execution core inference-rules, part 1.\todo[inline]{AL: this stuff is sticking out on the right side. I think we are going to be in trouble with this later.}}
  \label{nfm2023:fig:infer_rules_p1}
\end{figure}

\subsection{Inference rules}
The core set of inference rules
in Fig.~\ref{nfm2023:fig:infer_rules_p1}~and~Fig.~\ref{nfm2023:fig:infer_rules_p2}
form the foundation of the symbolic execution.
Each rule is a core\todo{AL: should we say 'basic' operation?} operation that an execution engine can carry out while guaranteeing sound results.

\smartparagraph{Rule \opref{\opstep}}
This is the base rule to obtain a  progress structure from a single step of the symbolic semantics.
Because the concrete transition relation and the function $\EsymbexecF{\symb{\state}}$ are total, the simulation of Theorem~\ref{nfm2023:thm:sym_step_sound} implies that the corresponding progress structure is sound.

\smartparagraph{Rules \opref{\opcase{}}, \opref{\opinf{}}, \opref{\opren{}}, and \opref{\opfreesymbren{}}}
While the symbolic semantics can already generate multiple states for individual instructions (e.g., for conditional branches), the rule \opref{\opcase{}} allows further splitting a target state by a path condition conjunct.
This can be useful to analyze different cases of the same execution path individually (e.g., one representing the case
that two registers point to different memory locations and the one representing memory aliasing).
The rule \opref{\opinf{}} allows dropping of infeasible paths, and \opref{\opren{}} allows straightforward renaming of symbols.
Rule \opref{\opfreesymbren{}} allows renaming of free symbols specifically, under more relaxed assumptions.

\smartparagraph{Rule \opref{\opsubst}}
This rule allows to derive specialized progress structures
from general ones, by substituting single symbols with symbolic expressions.
This is needed for reusing previous computations of binary fragments, e.g., the same code executing multiple times on several paths.
By requiring that the substituted symbol is bound, the rule prevents the substitution of free symbols, to avoid missing concrete executions.
Additionally, the symbols of the symbolic expression must not appear as free symbols, because they would become bound symbols otherwise, which may also lead to missing concrete executions.

\smartparagraph{Rule \opref{\opfreshif}}
This rule combines two purposes: simplifying a  symbolic expression of a target state by replacing it with an equivalent one and also introducing a fresh symbol in the new symbolic expression.

Simplification requires the expressions to be equivalent under the path condition of the state that is being changed.
For example, if the path condition entails $\Bsymsval_b > 0$, then we may use this rule to replace $\Bsymsval_a * \Bsymsval_b > 0$ with $\Bsymsval_a > 0$ in the store.\todo{RG: consistency: I've seen usage of both state and store. AL: should not be used as the same. state includes store and path condition. if we refer to state, then it applies to both. if we refer to store, it applies only to the store}
This rule is particularly important for memory operations of binary code, by allowing for the simplification of store and load expressions under the path condition, as 
identified and applied in previous work \cite{farinier2018arrays, Daniel2020, Cha2012}.
In this way, load expressions can be matched with store expressions if their addresses agree in all interpretations, or bypass them if the accessed memory range is always completely disjoint.
Similarly, store expressions can be removed if they are superseded by stores to the same addresses.

\begin{figure*}[t]
 \centering
  \begin{prooftree}
  	\hypo{
  	  \begin{array}{c}
  	    \vdash \HLsymbexec{\lbls}{\symb{\state}}{\HLsymbexecStruct \cup \{(\lbl, \symb{\Benv}, \bspcond)\}}\\
  	    \symb{\Benv}(\Bvarsval) = \symb{\BVval}
  	  \end{array}}
  	\hypo{
  	  \begin{array}{c}
  	    \forall \Bsyminterp.\ \BsyminterpPHolds{\Bsyminterp}{\bspcond \wedge \Bsymsval = \symb{\BVval}'' \Rightarrow \symb{\BVval} =  \symb{\BVval}'}\\
  	    \Bsymsval \not{\in} \SymbolsOf{\symb{\state}} \cup \SymbolsOf{(\lbl, \symb{\Benv}, \bspcond)}\\
            \SymbolsOf{\symb{\BVval}''} \subseteq \SymbolsOf{(\lbl, \symb{\Benv}, \bspcond)}
  	  \end{array}}
    \infer2[\optabref{\opfreshif}]{\vdash \HLsymbexec{\lbls}{\symb{\state}}{\HLsymbexecStruct \cup \{(\lbl, \Benvupd{\symb{\Benv}}{\Bvarsval}{\symb{\BVval}'}, \bspcond \wedge \Bsymsval = \symb{\BVval}'')\}}}
  \end{prooftree}
\\\vspace{1.0em}
  \begin{prooftree}
  	\hypo{\begin{array}{l}
  	  \symb{\state}_1 \Rightarrow \symb{\state}_1' \\
  	  \symb{\state}_2 \Rightarrow \symb{\state}_2'
  	\end{array}}
  	\hypo{\begin{array}{l}
  	  \vdash \HLsymbexec{\lbls}{\symb{\state}_1'}{\HLsymbexecStruct \cup \{\symb{\state}_2\}} \\
  	  \SymbolsOf{\symb{\state}_1} \cap \freshsym{\HLsymbexec{\lbls}{\symb{\state}_1'}{\HLsymbexecStruct \cup \{\symb{\state}_2\}}} = \emptyset
  	\end{array}}
    \infer2[\optabref{\opconseq}]{\vdash \HLsymbexec{\lbls}{\symb{\state}_1}{\HLsymbexecStruct \cup \{\symb{\state}_2'\}}}
  \end{prooftree}
\\\vspace{1.0em}
  \begin{prooftree}
  	\hypo{\begin{array}{l}
  	  \vdash \HLsymbexec{\lbls}{\symb{\state}}{\HLsymbexecStruct \cup \{(\lbl, \symb{\Benv}, \bspcond)\}}
  	\end{array}}
  	\hypo{\begin{array}{l}
  	    \forall \Bsyminterp.\ \BsyminterpPHolds{\Bsyminterp}{\bspath{\symb{\state}} \Rightarrow \bspcond'} \\
  	  \SymbolsOf{\bspcond'} \subseteq \SymbolsOf{\bspath{\symb{\state}}}
  	\end{array}}
    \infer2[\optabref{\optransf}]{\vdash \HLsymbexec{\lbls}{\symb{\state}}{\HLsymbexecStruct \cup \{(\lbl, \symb{\Benv}, \bspcond \wedge \bspcond')\}}}
  \end{prooftree}
\\\vspace{1.0em}
  \begin{prooftree}
  	\hypo{\vdash \HLsymbexec{\lbls_A}{\symb{\state}_A}{\HLsymbexecStruct_A}}
  	\hypo{\vdash \HLsymbexec{\lbls_B}{\symb{\state}_B}{\HLsymbexecStruct_B}}
  	\hypo{\SymbolsOf{\symb{\state}_A} \cap \freshsym{\HLsymbexec{\lbls_B}{\symb{\state}_B}{\HLsymbexecStruct_B}} = \emptyset}
    \infer3[\optabref{\opseq}]{\vdash \HLsymbexec{\lbls_A \cup \lbls_B}{\symb{\state}_A}{\left(\HLsymbexecStruct_A \setminus \{\symb{\state}_B\}\right) \cup \HLsymbexecStruct_B}}
  \end{prooftree}
  \caption{Overview of the symbolic-execution core inference-rules, part 2.}
  \label{nfm2023:fig:infer_rules_p2}
\end{figure*}

The introduction of a fresh symbol $\Bsymsval$ requires that it does not occur in the source state, and also not in the target state where it is introduced.
Otherwise, this could interfere with other usages of this symbol as per the soundness property of the progress structure.
To avoid unnecessary overapproximation, the path condition of the modified state is extended with a constraint on $\Bsymsval$.
Intuitively, it is fixed to the value of the original symbolic expression that it replaces in the modified target state, to retain exactly the feasible set of state concretizations and not add any overapproximation in the process.
%
For example, executing line 8 of Fig.~\ref{fig:modexp_imp_cfg} with \opref{\opstep{}} produces the sound progress structure\todo{AL: we use the same example statement above in the progress structure soundness introduction. should we try to save some space? or try to keep the symbol naming the same?} 
$\HLsymbexec{\{8\}}{(8, \symbbenvinit, T)}{\{(9, \Benvupd{\symbbenvinit}{\registername{R1}}{\symb{\BVval}}, T)\}}$,
where $\symb{\BVval} = (\registername{R1} * \registername{R1}) \% \registername{R2}$.
Using \opref{\opfreshif{}}, we can introduce $\Bsymsval$ in place of $\symb{\BVval}$ and obtain
$\HLsymbexec{\{8\}}{(8, \symbbenvinit, T)}{\{(9, \Benvupd{\symbbenvinit}{\registername{R1}}{\Bsymsval}, \Bsymsval = \symb{\BVval})\}}$.

\smartparagraph{Rule \opref{\opconseq}}
This rule allows to strengthen the path condition of the source state and to weaken the path condition of a target state.
Free symbols of the structure
$\HLsymbexec{\lbls}{\symb{\state}_1'}{\HLsymbexecStruct \cup \{\symb{\state}_2\}}$
must not occur in
$\symb{\state}_1$
to avoid capturing them.

We say that $\symb{\state}$ is weaker than $\symb{\state}'$, written as $\symb{\state} \Rightarrow \symb{\state}'$, if the two states have the same store and program counter and the path condition of $\symb{\state}$ implies the path condition of $\symb{\state}'$, i.e.,
$\symb{\state}  = (\lbl, \symb{\Benv}, \bspcond) \wedge \symb{\state}' = (\lbl, \symb{\Benv}, \bspcond') \wedge \forall \Bsyminterp.\ \BsyminterpPHolds{\Bsyminterp}{\bspcond \Rightarrow \bspcond'}$.
It is obvious that $\symb{\state} \Rightarrow \symb{\state}'$ implies that every state $\state$ matched by $\symb{\state}$ is also matched by $\symb{\state}'$.
%

Applying \opref{\opconseq{}} to the example before allows removing the constraint on $\Bsymsval$ from the path condition of the final state, weakening it, and obtaining
$\HLsymbexec{\{8\}}{(8, \symbbenvinit, T)}{\{(9, \Benvupd{\symbbenvinit}{\registername{R1}}{\Bsymsval}, T)\}}$.
This results in overapproximation, since the value of \registername{R1} is now unconstrained.

\smartparagraph{Rule \opref{\optransf}}
This rule is complementary to \opref{\opconseq{}} and it allows for strengthening the target path conditions based on the source path condition.
This is permitted because the interpretation $\Bsyminterp$ in Definition~\ref{nfm2023:def:sym_exec_soundness} is extended when loosely matching target states, and therefore the initial path condition is also satisfied in all target states.

\smartparagraph{Rule \opref{\opseq}}
Finally, this rule allows the sequential composition of two symbolic executions.
This operation implicitly widens the code fragment scope by merging the two label sets.
Applying this operation only makes sense in case $\symb{\state}_B$ occurs in $\HLsymbexecStruct_A$, which also does not introduce overapproximation.
We require that there is no free symbol in the second execution ($B$) that is bound in the first execution ($A$).
Otherwise, the result may miss concrete executions.

To continue with the example from before, we need to first execute line 9 with \opref{\opstep{}} to obtain
$\HLsymbexec{\{9\}}{(9, \Benvupd{\symbbenvinit}{\registername{R1}}{\Bsymsval}, T)}{\{(10, \Benvupd{\Benvupd{\symbbenvinit}{\registername{R1}}{\Bsymsval}}{\registername{R3}}{\Bsymsval_{\registername{R3}}+1}, T)\}}$.
The sequential composition of the two progress structures renders 
$\HLsymbexec{\{8,9\}}{(8, \symbbenvinit, T)}{\{(10, \Benvupd{\Benvupd{\symbbenvinit}{\registername{R1}}{\Bsymsval}}{\registername{R3}}{\Bsymsval_{\registername{R3}}+1}, T)\}}$, which is allowed because the second progress structure has no free symbols.

\subsection{Symbolic execution example}
\label{nfm2023:app:symexec_app}
We illustrate our rules on the modular exponentiation of Fig.~\ref{fig:modexp_imp_cfg}. We focus on using symbolic execution to establish control-flow graph integrity, i.e., verifying that the link register \registername{R3} is correctly restored. We use this example to show how overapproximation helps scalability: it allows us to abstract from values irrelevant to control flow while still proving its integrity. A naïve executor that only applies \opref{\opstep} and \opref{\opseq} would quickly become inefficient due to branching and expression growth.

We start analysing the loop body. Starting from path condition $\bspcond_0$, executing line 6 with \opref{\opstep} yields this progress structure for the conditional branch:
$
\HLsymbexec{6}
{(6,\symbbenvinit, \bspcond_0)}{
\{
\symb{\state}_{T},
\symb{\state}_{F}
\}
}
$,
where $\symb{\state}_{T} =
(7,\symbbenvinit, \bspcond_0 \land (\Bsymsval_{\registername{R0}} >> \Bsymsval_1) \& 1) $ and $\symb{\state}_{F}
(8,\symbbenvinit, \bspcond_0 \land \neg ((\Bsymsval_{\registername{R0}} >> \Bsymsval_1) \& 1))
$.
The two final states are produced directly by applying the symbolic semantics for individual instructions, rather than manually identifying the branch conditions and applying \opref{\opcase{}}.

The two paths can be analysed individually, starting from the corresponding state produced above, simply using  \opref{\opstep} and \opref{\opseq}. This results in 
$
\HLsymbexec{[7 \dots 9]}
{\symb{\state}_{T}}
{\symb{\state}_{1}}
$,
where $\symb{\state}_{1} =
(10, \symb{\Benv_{1}},
\bspcond_0 \land (\Bsymsval_{\registername{R0}} >> \Bsymsval_1) \& 1)$
and
$\symb{\Benv_{1}} =
\Benvupd
    {
        \Benvupd{
            \Benvupd{\symbbenvinit}
            {\registername{R3}}{\Bsymsval_1 + 1}
        }
        {\registername{R1}}{\Bsymsval_{\registername{R1}} * \Bsymsval_{\registername{R1}} \% \Bsymsval_{\registername{R2}}}
    }
  {\registername{M}}
  {\Best
    {\Bsymsval_{\registername{M}}}
    {\Bsymsval_{\registername{SP}}}
    {\Beld
        {\Bsymsval_{\registername{M}}}
        {\Bsymsval_{\registername{SP}}}
    * \Bsymsval_{\registername{R1}} \% \Bsymsval_{\registername{R2}}
    }
 }
$, and 
$
\HLsymbexec{[7 \dots 9]}
{\symb{\state}_{F}}
{\symb{\state}_{2}}
$,
where $
{\symb{\state}_{2}} = 
{(10, \symb{\Benv_{2}},
\bspcond_0 \land \neg (\Bsymsval_{\registername{R0}} >> \Bsymsval_1) \& 1)} 
$ and
$\symb{\Benv_{2}} = \Benvupd{
            \Benvupd{\symbbenvinit}
            {\registername{R3}}{\Bsymsval_1 + 1}
        }
        {\registername{R1}}{\Bsymsval_{\registername{R1}} * \Bsymsval_{\registername{R1}} \% \Bsymsval_{\registername{R2}}}$.
These two progress structures one by one with the one obtained from line 6 via \opref{\opseq} renders:
$
\HLsymbexec{[6 \dots 9]}{(6,\symbbenvinit, \bspcond_0)}{
\{
\symb{\state}_{1} ,\symb{\state}_{2}
\}
}
$.

Overapproximating the behavior of the loop reduces the complexity of the analysis.
For instance, when establishing control flow integrity and termination,
the values 
of $\registername{R1}$ and the stack variable where $r$ is stored are irrelevant.
Overapproximating the former allows us to avoid duplicating the size of the expression for $\registername{R1}$ at each
loop iteration (due to the double reference to the previous value of $\registername{R1}$).
Overapproximating the latter allows us to merge the two symbolic states resulting from the loop execution.
In both cases, the approach is similar.

\todo[inline]{RG: This description is incomplete and not correct. Must check with Andreas}
For the memory operation,
we first transform $\symb{\state}_{2}$ to an identical one \revision{AL}{R3.P12}{}{}{using \opfreshif} by explicitly storing the current memory value at $\Bsymsval_{\registername{SP}}$:
$(10,
\Benvupd
    {\symb{\Benv_{2}}}
  {\registername{M}}
  {\Best
    {\Bsymsval_{\registername{M}}}
    {\Bsymsval_{\registername{SP}}}
    {\Beld
        {\Bsymsval_{\registername{M}}}
        {\Bsymsval_{\registername{SP}}}
    }
 }
,$
$\bspcond_0 \land \neg (\Bsymsval_{\registername{R0}} >> \Bsymsval_1) \& 1)
$.
Then for both states, we use \opref{\opfreshif} to introduce the fresh symbol $\Bsymsval'$ for $\Beld
        {\Bsymsval_{\registername{M}}}
        {\Bsymsval_{\registername{SP}}}
     * \Bsymsval_{\registername{R1}} \% \Bsymsval_{\registername{R2}}
$ and $\Beld
        {\Bsymsval_{\registername{M}}}
        {\Bsymsval_{\registername{SP}}}
    $ respectively to make the two symbolic stores identical.
Finally, we remove the equality introduced in the path condition and drop the branch condition conjuncts introduced at line 6 with \opref{\opconseq},
which allows to merge the two states that are now identical and obtain the progress structure:
$
\HLsymbexec{[6 \dots 9]}
{(6,\symbbenvinit, \bspcond_0)}
{(10,
\Benvupd
    {
        \Benvupd{
            \Benvupd{\symbbenvinit}
            {\registername{R3}}{\Bsymsval_1 + 1}
        }
        {\registername{R1}}{\Bsymsval_{\registername{R1}} * \Bsymsval_{\registername{R1}} \% \Bsymsval_{\registername{R2}}}
    }
  {\registername{M}}
  {\Best
    {\Bsymsval_{\registername{M}}}
    {\Bsymsval_{\registername{SP}}}
    {\Bsymsval'}
 },
\bspcond_0)}
$.

Merging two states to avoid path explosion is a common technique~\cite{Kuznetsov2012} and similar to widening for reaching fixed points in abstract interpretation~\cite{cousot1996abstract}.
A similar approach can be used to introduce a new symbol $\Bsymsval''$ that allows to abstract from the details of the expression computed for \registername{R1}:
$
\HLsymbexec{[6 \dots 9]}
{(6,\symbbenvinit, \bspcond_0)}
{(10,
\Benvupd
    {
        \Benvupd{
            \Benvupd{\symbbenvinit}
            {\registername{R3}}{\Bsymsval_1 + 1}
        }
        {\registername{R1}}{\Bsymsval''}
    }
  {\registername{M}}
  {\Best
    {\Bsymsval_{\registername{M}}}
    {\Bsymsval_{\registername{SP}}}
    {\Bsymsval'}
 },
\bspcond_0)}
$.

The progress structure for the loop body can be used when creating a progress structure for the whole example function.
The process for this is sequential execution from the first line, greedily instantiating the loop body progress structure whenever possible.
Instantiation of progress structures requires instantiation of symbols of the state via \opref{\opsubst{}} and massaging of expressions (like reordering store operations in symbolic memory) via \opref{\opfreshif} and \opref{\opren{}}.
Additionally, after each instantiation and subsequent composition with \opref{\opseq}, it is necessary to remove the infeasible branch of the loop body progress structure with \opref{\opinf} and drop the introduced conjunct with \opref{\opconseq}.
Using \opref{\opfreshif}, we remove store operations that have the same symbolic address ($\Bsymsval_{\registername{SP}} - 4$), thus maintaining an overall smaller memory expression.
This is a standard simplification enabled by array axiomatization~\cite{farinier2018arrays, Daniel2020, Cha2012}.
After completing all operations, we finally obtain the overall progress structure:
\[
\begin{array}{c}
\HLsymbexecLeft{\{1..14\}}{
(1, \symbbenvinit, \bspcond_0)
}
\{
(15,
\symbbenvinit
\left[
\begin{array}{c}
\registername{SP} \mapsto \Bsymsval_{\registername{SP}},\ 
\registername{R0} \mapsto \Bsymsval_a,\ 
\registername{R1} \mapsto \Bsymsval_b,\ 
\registername{R3} \mapsto \Bsymsval_{\registername{R3}}
\\
\registername{M} \mapsto \Best{\Best{\Bsymsval_{\registername{M}}}
  {\Bsymsval_{\registername{SP}}}{\Bsymsval_{\registername{R3}}}}
  {\Bsymsval_{\registername{SP}} - 4}{\Bsymsval_a}
\end{array}
\right]
, \bspcond_0)
\}
\end{array}
\]

This progress structure has only one reachable state, which
implies guaranteed termination in a matching state at line 15, at which
the registers \registername{SP} and \registername{R3} are restored to their initial values.

\section{Implementation}
\label{nfm2023:sec:impl}

This section gives an overview of our implementation in the HOL4 theorem prover of our symbolic execution theory and its instantiation for BIR. 

\begin{figure}[t]
\centering
\includegraphics[width=0.9\textwidth]{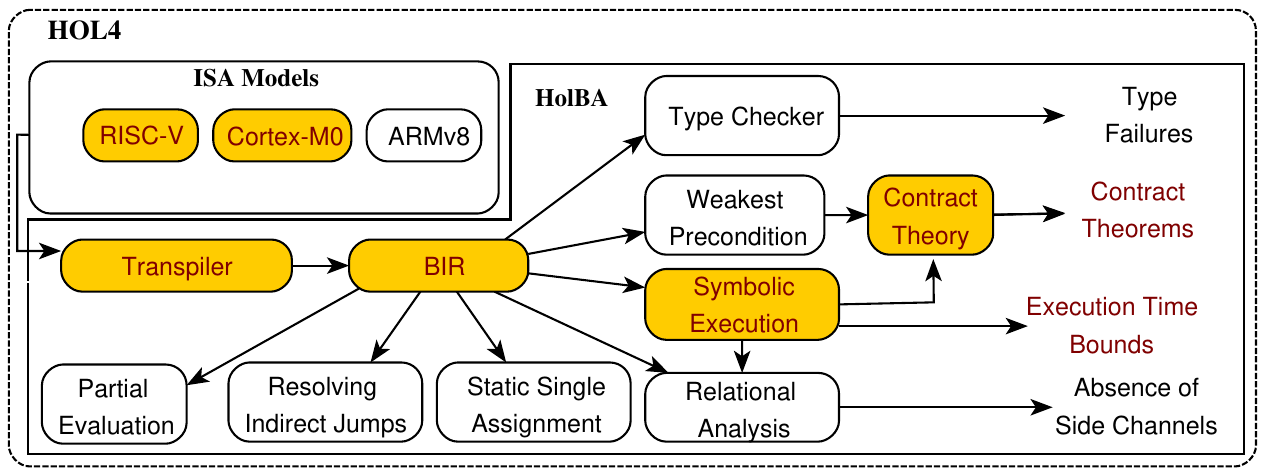}
\caption{Architecture of HolBA, a HOL4-based binary analysis library. Components and outputs relevant to \toolchain are highlighted.} 
\vspace{-1em}
\label{fig:holba}
\end{figure}

We build on HolBA, a library for binary analysis implemented in Standard ML (SML) using the HOL4 theorem prover~\cite{hol4}.
HolBA follows the architecture of traditional binary analysis tools such as the Binary Analysis Platform (BAP)~\cite{bap} in that it \emph{transpiles} machine code to a single intermediate language (BIR for HolBA) in which analyses are performed. In contrast to BAP, HolBA builds directly on formal ISA models as defined in their L3 specifications~\cite{Fox2015,FoxL3modelsweb}, and produces HOL4 \emph{lifting} theorems linking machine code behavior with transpiled BIR behavior via simulation relations. This significantly reduces the trusted computing base (TCB) of HolBA compared to similar tools. Fig.~\ref{fig:holba} illustrates the organization of HolBA and the relations between its components.

Our extension, \toolchain, adds to HolBA (1) a standalone HOL4 theory on symbolic execution and progress structures, (2) an instantiation of this theory for BIR, (3) general SML procedures automating proofs of progress structure soundness and program properties, (4) an integration of these to perform automatic verification of (i) contracts for RISC-V programs and (ii) execution time bounds for ARM Cortex-M0 programs. 
\todo[inline]{AL: I think this paragraph (part above) carries implicit information (eg, instantiation of rules could also be done otherwise: without HOL4 etc), but in the interest of space, it seems like a good choice}
Thanks to the embedding of HOL4 in SML, each \toolchain verification task can be manifested as a custom SML program that reads files such as disassembly and then generates and checks the relevant HOL4 definitions and theorems.


\subsection{Results formalized and proved in HOL4}

Our standalone theory of symbolic execution in HOL4 encodes the definitions and proofs of all rules from Sec.~\ref{sec:symexec} using predicate sets and inductive relations. 
To integrate this theory with BIR as represented in HolBA, we encoded the symbolic semantics as described in Sec.~\ref{nfm2023:sec:symsem} and proved Theorem~\ref{nfm2023:thm:sym_step_sound} in HOL4, establishing that the semantics is sound with respect to the existing concrete BIR semantics.
The definitions and proofs of the symbolic execution rules consist of around 16,000 lines of HOL4 (SML) code and took 8 person months to develop. 

\subsection{Symbolic executors and progress structures in \toolchain}
Our approach to practical symbolic execution of BIR is to provide a framework inside HOL4 and HolBA where symbolic executors with different strategies are expressed as proof-producing SML procedures. Here, \emph{strategy} refers to how the executor composes underlying execution blocks when run, and \emph{proof-producing} refers to how a procedure produces a HOL4 theorem when (or if) it terminates\todo{AL: i am wondering if 'how' and 'produces' is explaining something or is clear here. also, i am specifically wondering if producing a theorem makes clear to a reader that it is a verified proof/machine-checked. maybe some edit i did earlier is enough? just saying because we had such a review feedback earlier}. In executors, such theorems are witnesses that a specific progress structure (``summary'') has the required properties\todo{AL: i am thinking that required properties here is too abstract. i think the main point is that the progress structures are sound (because we defined that as main property) and then we derive other properties, independent of the progress structures, about executions}. In a program verification workflow, progress structure theorems obtained from executors are decomposed and reused in other theorems, e.g., in Hoare-style contract proofs as described in Sec.~\ref{sec:riscv-application}.

In practice, our framework is a high-level abstraction layer (SML API) for leveraging symbolic execution rules. For a selection of rules, we provide parameterized execution blocks that perform one primitive symbolic operation and return a HOL4 theorem. When running an executor, such parameterized blocks are composed and interleaved with domain-specific post-processing steps. For example, one may apply the infeasible rule to prune branches, or simplify expressions for constant propagation and memory accesses\todo{AL: i am wondering about the choice of term 'post-processing' here. it seems you contrast this to the primitive symbolic operation, but i think this is not really post the other. just saying it might cause some confusion if we don't say what exactly the processing is. maybe it is enough to choose a name that doesn't raise the question of a dual and is also able to capture the processing? maybe we just call those complex combinations of primitive operations (the ones closest to the the rules of section 4), where some happen to be executing from-to with instantiation and others are more post-execution-processing, and yet others may do even some program specific stuff like forgetting expressions to be able to merge a loop body or whatever}. ISAs may require different approaches due to specific uses of expression combinations. We developed symbolic executors for RISC-V and ARM Cortex-M0 with this in mind.



\subsection{Practical evaluation of symbolic execution using HolBA-SE}
\label{nfm2023:sec:impl:eval}
For application in practical automated verification tasks related to binary code, a symbolic executor needs to be fast and scalable enough to handle hundreds of instructions with complex control flow. To achieve this, we developed optimized executors for RISC-V and ARM Cortex-M0 and integrated them with the SMT solver Z3~\cite{z3}, letting it discharge conditions that are BIR expressions, constant program memory lookups and symbolic value concretizations. This approach increases the TCB, but we believe the trade-off is attractive since the performance boost is considerable and the scope of discharged conditions is restricted. 

\begin{table}[ht]
\centering
\begin{tabular}{l|l|r|r}
\textbf{Program} & \textbf{Description} & \textbf{\#Instrs} & \ \ \textbf{Time}\\ 
\hline \hline
aes-unopt   & AES cipher, -O0 optimization & 310                      & 71.38 s \\
\hline
aes         & AES cipher, -O1 optimization & 130                     & 6.84 s \\
\hline
chacha20     & ChaCha20 cipher & 177                      & 46.28 s \\
\hline
incr         & increment variable & 1                    & 0.04 s\\ 
\hline
isqrt        & integer square root & 6                  & 0.12 s \\
\hline
kernel-trap  & S3K kernel trap routines & 79                    & 9.21 s \\
\hline
mod2          & modulo two computation  & 1                          & 0.09 s \\
\hline
modexp       & modular exponentiation & 10 & 0.26 s\\
\hline
motor\_set   & nested calls and branching & 66          & 3.94 s \\
\hline
swap          & swap memory location   & 5               & 0.19 s \\
\end{tabular}
\vspace{0.3cm}
\caption{RISC-V programs and their symbolic execution running times. The number of instructions are for instructions processed during symbolic execution.} \label{tbl:symbexec}
\vspace{-0.5cm}
\end{table}

To investigate performance and scalability in practice, we applied our symbolic executor for RISC-V to a collection of disassembled binaries and measured the time to generate a progress structure theorem; the results are shown in Table~\ref{tbl:symbexec}. 
All experiments were run on an Intel Core i5-9600K with 32GB RAM.
Per the table, building HOL4 progress structure theorems for BIR programs corresponding to disassembly with hundreds of instructions generally takes minutes at most. Time increases with both the number of BIR instructions symbolically executed and the \textit{store complexity}, which refers to how much information needs to be stored in the symbolic state. In ciphers where all output bits depend on all input bits, expressions may be replicated many times in symbolic memory.
\section{Application: Automated Verification of Functional Properties of RISC-V Binaries}
\label{sec:riscv-application}
We automated functional verification in \toolchain for RISC-V binaries by integrating symbolic execution with previous work on binary lifting to BIR~\cite{Lindner2019} and a theory of Hoare-style binary contracts~\cite{Lundberg2020}.
The result is a workflow with the steps below as illustrated in Fig.~\ref{fig:workflow},
supporting the RV64G instruction set variant of RISC-V, a collection of uncompressed general 64-bit instructions~\cite{riscv}. As with most theorem prover based verification, the workflow is \emph{incomplete} in the sense that verification steps can fail without specifications being false.

\begin{figure}[t]
\centering
\begin{tikzpicture}
\node[rectangle,draw=black,rounded corners, minimum width=1.8cm, align=center, minimum height=1cm,dashed] (src) {Source\\code};
\node[rectangle,draw=black,rounded corners, minimum width=1.8cm, minimum height=1cm, align=center,right of=src, node distance=2.6cm] (bin) {Binary};
\node[rectangle,draw=black,rounded corners, minimum width=1.8cm, minimum height=1cm, align=center,right of=bin, node distance=2.6cm] (dis) {Disassembly};
\node[rectangle,draw=black,rounded corners, minimum width=1.8cm, minimum height=1cm, align=center,below of=dis, node distance=1.7cm] (riscvspec) {RISC-V\\spec.};
\node[rectangle,draw=black,rounded corners, minimum width=1.8cm, minimum height=1cm, align=center,left of=riscvspec, node distance=2.6cm] (birspec) {BIR spec. \&\\program};
\node[rectangle,draw=black,rounded corners, minimum width=1.8cm, minimum height=1cm, align=center,left of=birspec, node distance=2.6cm] (struct) {Progress\\structure};
\node[rectangle,draw=black,rounded corners, minimum width=1.8cm, minimum height=1cm, align=center,left of=struct, node distance=2.6cm] (birc) {BIR\\contract};
\node[rectangle,draw=black,rounded corners, minimum width=1.8cm, minimum height=1cm, align=center,above of=birc, node distance=1.7cm] (riscvc) {RISC-V\\contract};

\draw[->, very thick, >=stealth] (src) -- (bin) node[midway, circle, draw, thick, inner sep=2pt,yshift=0.3cm] {0};
\draw[->, thick, >=stealth] (bin) -- (dis) node[midway, circle, draw, thick, inner sep=2pt,yshift=0.3cm] {1};
\draw[->, thick, >=stealth] (dis) -- (riscvspec) node[midway, circle, draw, thick, inner sep=2pt,xshift=0.3cm] {3};
\draw[->, thick, >=stealth] (dis) -- (birspec) node[midway, circle, draw, thick, inner sep=2pt,xshift=0.2cm,yshift=-0.25cm] {2};
\draw[->, thick, >=stealth] (riscvspec) -- (birspec) node[midway, circle, draw, thick, inner sep=2pt,yshift=-0.3cm] {4};
\draw[->, thick, >=stealth] (birspec) -- (struct) node[midway, circle, draw, thick, inner sep=2pt,yshift=-0.3cm] {5};
\draw[->, thick, >=stealth] (struct) -- (birc) node[midway, circle, draw, thick, inner sep=2pt,yshift=-0.3cm] {6};
\draw[->, thick, >=stealth] (birc) -- (riscvc) node[midway, circle, draw, thick, inner sep=2pt,xshift=0.3cm] {7};
\draw[->, thick,bend left=20, >=stealth] (dis) to (riscvc);
\draw[->, thick,bend right=20, >=stealth] (birspec) to (birc);
\end{tikzpicture}
\caption{Binary analysis workflow using \toolchain; source code is optional.} 
\label{fig:workflow}
\end{figure}

\begin{enumerate}[start=0]
\item \textbf{Compilation:} Compile a source program to a RV64G binary. This step is \emph{optional}, but adjustments to compilation can affect verification. 
\item \textbf{Disassembly:} Disassemble the RV64G binary. 
\item \textbf{Lifting:} Transpile the disassembly to BIR and generate a HOL4 lifting theorem connecting the BIR program to corresponding RV64G binary code.
\item \textbf{RISC-V specification:} Specify program boundaries inside disassembly and define pre- and post-conditions using the L3 model of RISC-V.
\item \textbf{BIR specification:} Translate RISC-V pre-conditions and post-conditions to BIR expressions and prove equivalence in HOL4.
\item \textbf{Symbolic execution:} Symbolically execute the BIR program in HOL4 according to the program boundaries and obtain a progress structure theorem.
\item \textbf{Symbolic transfer:} Prove a HOL4 contract theorem for the BIR program using the pre- and post-conditions via the progress structure theorem. 
\item \textbf{Backlifting:} Prove a RISC-V contract theorem by way of the BIR contract theorem and the lifting theorem.
\end{enumerate}
All steps except 3 and 4 (writing and translating specifications) are mostly or fully automated. Steps 5, 6, and 7 are novel contributions.

\subsection{Contract proof automation using symbolic execution}
Steps 5 and 6 in the workflow in Fig.~\ref{fig:workflow} crucially rely on sound progress structures produced by symbolic execution to prove Hoare-style contracts for BIR programs. The idea is that since our approach allows overapproximation, 
we can use a progress structure
to establish a postcondition $Q$ starting from a precondition $P$, but we cannot in general guarantee that it is the strongest one.
The following theorem connects progress structures and Hoare-style contracts.



%

\begin{theorem}
\label{nfm2023:thm:transf_contr}
Given $\vdash \HLsymbexec{\lbls}{\symb{\state}}{\HLsymbexecStruct}$,
let $\bar{\lbls}$ be the complement of $\lbls$. Then, if
\begin{itemize}
    \item for every $\state$ if $\pc{\state} = \pc{\symb{\state}} $ and $P(\state)$ then there exists $\Bsyminterp$ such that $\symmatch{\symb{\state}}{\Bsyminterp}{\state}$
    \item for every $\symmatch{\symb{\state}}{\Bsyminterp}{\state}$, $\symb{\state}' \in \HLsymbexecStruct$, and $\symmatchext{\symb{\state}'}{\Bsyminterp}{\state'}$ if $P(\state)$ then $Q(\state')$
    \item for every $\symb{\state}' \in \HLsymbexecStruct$, $\pc{\symb{\state}'} \in \bar{\lbls}$ holds,
\end{itemize}
we have that for every state such that $\pc{\state} = \lbl$ and $P(\state)$ there is a state $\state'$ and an $n > 0$ such that $\transitionsf{\state}{n}{\lbls}{\state'}$, $\pc{\state'} \in \bar{\lbls}$,
 and $Q(\state')$.
\end{theorem}

\subsection{Workflow example}

To explain the steps in our workflow, we use the example defined in Fig.~\ref{fig:ex-prog}, which includes a simple program \texttt{incr} that increments its argument by one.


\begin{figure}[t]
\centering
\adjustbox{varwidth=\linewidth,scale=0.8}{%
\lstset{language=C,
    basicstyle=\ttfamily,
    keywordstyle=\bfseries,
    showstringspaces=false,
    frame=single,
    morekeywords={include, printf}
}
\begin{lstlisting}
#include <stdint.h>
uint64_t incr(uint64_t i) { return i + 1; }
\end{lstlisting}
\lstset{
    language={[RISCV]Assembler},  
    basicstyle=\ttfamily,
    keywordstyle=\bfseries,
    showstringspaces=false,
    frame=single,
}
\vspace{-0.5em}
\begin{lstlisting}
incr:     file format elf64-littleriscv
Disassembly of section .text:
0000000000010488 <incr>:
   10488:	00150513          	addi	a0,a0,1
   1048c:	00008067          	ret
\end{lstlisting}
\lstset{
    language=SML,  
    basicstyle=\ttfamily,
    keywordstyle=\bfseries,
    showstringspaces=false,
    frame=single,
}
\vspace{-0.5em}
\begin{lstlisting}
Definition riscv_incr_pre_def: (* RISC-V precondition *)
 riscv_incr_pre (pre_x10:word64) (m:riscv_state) : bool =
 (m.c_gpr m.procID 10w = pre_x10)
End
Definition riscv_incr_post_def: (* RISC-V postcondition *)
 riscv_incr_post (pre_x10:word64) (m:riscv_state) : bool =
 (m.c_gpr m.procID 10w = pre_x10 + 1w)
End
\end{lstlisting}\vspace{-0.5em}
\begin{lstlisting}
Theorem riscv_cont_incr: (* RISC-V contract *)
 riscv_cont incr_progbin 0x10488w {0x1048cw}
  (riscv_incr_pre pre_x10) (riscv_incr_post pre_x10)
\end{lstlisting}
}
\caption{Example verified program that increments a 64-bit unsigned integer represented in C (top), corresponding RV64G disassembly (below top), and RISC-V specification (below disassembly) and contract theorem in HOL4 (bottom).} \label{fig:ex-prog}
\end{figure}


%
For step 0, \emph{compilation}, our toolchain and workflow only require a RV64G binary. However, we expect many users to perform binary analysis on programs for which they have source code, enabling control over optimizations which can affect binary size. Fig.~\ref{fig:ex-prog} (top) shows the source for the example.
%
%
In step 1, we \emph{disassemble} RV64G binaries using  \texttt{gnu-objdump} for further processing. The middle box in Fig.~\ref{fig:ex-prog} shows our example's disassembly. This step is not strictly necessary, since lifting can be performed directly on the binary, but we found disassembly useful as a reference for specification. 
%
%
For step 2, \emph{lifting}, \toolchain parses the disassembly using specifications of program bounds (address of initial instruction and address after final instruction to lift) and translates the indicated input into a BIR program inside HOL4 along with a lifting theorem connecting input and output. 
%
%
For step 3, \emph{RISC-V specification}, we express binary contracts as pairs of pre- and postconditions in HOL4 using the \texttt{riscv} machine state type defined by the L3 model, as illustrated in the middle of Fig.~\ref{fig:ex-prog}, where \texttt{a0} is the standardized application binary interface (ABI) name of the register \texttt{x10}. Hence, the postcondition says that register \texttt{a0} is incremented by one.

For step 4, \emph{BIR specification}, we manually translate RISC-V pre- and postconditions to BIR to enable symbolic execution. 
%
Using the BIR specification, we \emph{symbolically execute} the BIR program and perform \emph{transfer} and \emph{backlifting} (steps 5, 6, and 7). We use the BIR precondition as a path condition and leverage the progress structure theorem to automatically prove a BIR contract via Theorem~2, and finally obtain a RISC-V contract, as shown in the lowermost parts of Fig.~\ref{fig:ex-prog}.
The contract primitive \texttt{riscv\_cont} was defined in earlier work~\cite{Lundberg2020} and uses the HOL4 representation of the increment program disassembly (\texttt{incr\_progbin}).

\subsection{Case studies}

To validate that \toolchain can be used to verify real-world RISC-V binaries, we performed case studies for two high-assurance system software domains: cryptography (ChaCha20 encryption) and operating systems (S3K trap functions).

\smartparagraph{ChaCha20 encryption}
ChaCha20 is a stream cipher that performs twenty rounds of the ChaCha algorithm as defined by Bernstein~\cite{Bernstein2008,rfc8439}. We obtained a ChaCha20 binary by compiling a reference implementation in C using GCC with optimization level O1.
We focused on specifying and verifying the byte encryption loop body of the ChaCha20 binary, taking inspiration from similar specifications in hacspec~\cite{bhargavan2018hacspec}. We abstractly define a ChaCha round in HOL4 as a transformation of 16 variables holding unsigned 32-bit words (array in the source code), built up from \emph{lines} that update two variables at a time using word addition, XOR (\texttt{??}), and left rotation (\lstinline[basicstyle=\ttfamily]{#<<~}):
\begin{lstlisting}[basicstyle=\ttfamily\small]
chacha_line a b c d s (m : word32 -> word32) =
 let m = (a =+ (m a) + (m b)) m in
 let m = (d =+ ((m d) ?? (m a)) #<<~ s) m in m
\end{lstlisting}
Our verification of byte encryption establishes the connection between this high-level specification in HOL4 (updating the ``memory function'' \texttt{m}) and how the loop body of the RISC-V binary behaves, as in functional translation validation. The verification process took around 4 minutes on the machine described in Sec.~\ref{nfm2023:sec:impl:eval}.

\smartparagraph{S3K trap functions}
S3K is an open source capability-based separation kernel targeting embedded RISC-V systems~\cite{s3k}. Practically, S3K manages a collection of 
\emph{user processes} and ensures that they adhere to given restrictions on execution time (slices), memory access, and interprocess communication. 
We specified and verified two key handwritten RISC-V assembly routines from S3K: \texttt{trap\_entry} and \texttt{trap\_exit}.
The routine \texttt{trap\_entry} is entered when the user process is interrupted or an exception is encountered. It loads a pointer to the process's process control block (PCB) from the special \texttt{mscratch} RISC-V register, and then stores the user process's context (general purpose registers) to the PCB.
The routine \texttt{trap\_exit} is entered when the kernel resumes a user process. It loads the user process's context from the PCB,  
replacing register values with the values from the PCB.
We obtained disassembly for the routines from the S3K author, consisting of 81 instructions. After adjusting lifting for special RISC-V instructions such as \texttt{csrrw}, we expressed routine contracts in terms of assignment of registers to memory locations (\texttt{trap\_entry}), and assignments of memory locations to registers (\texttt{trap\_exit}).
The S3K author validated our specifications, and we then translated them to BIR manually and performed the verification process separately for each routine, which took around 7 minutes in total on the machine described in Sec.~\ref{nfm2023:sec:impl:eval}.


\section{Application: Execution Time Bounds for Cortex-M0}
\label{nfm2023:sec:impl:exectime}
The ARM Cortex-M0 processor 
allows for predictable execution times, due to
lack of caches and a very short three-stage pipeline. 
According to the reference manual~\cite{CortexM0instrcycles}, ALU instructions take 1 cycle, while memory interactions take 1 extra cycle and control flow changes take 2 extra cycles. 
%
The Cortex-M0 formal model in HOL4~\cite{FoxL3modelsweb} maintains a clock cycle counter for this.
%
\todo{AL: do we need to mention somewhere that we have to maintain assumptions about the stack to be able to prove this? or not so important because it is generally needed for dealing with functions?}


\subsection{Representing execution time in BIR}
To accommodate this simple timing model, BIR programs increment a cycle counter variable $\registername{c}$ with the execution time needed by the instructions.
Throughout symbolic execution, the cycle counter \revision{AL}{R3.S7.1}{}{stays constant}{is an incrementing constant value} along each execution path.
When paths are merged, the cycle counter is approximated as an interval, where the bounds are encoded in the path condition, linked to the symbolic store by a free symbol.
For example, the cycle counter interval $[\Bsymsval_{\registername{c}} + 5, \Bsymsval_{\registername{c}} + 8]$ consists of $\BenvupdMap{\registername{c}}{\Bsymsval'}$ in the store and the path condition conjunct $\Bsymsval_{\registername{c}} + 5 \leq \Bsymsval' \leq \Bsymsval_{\registername{c}} + 8$.
The interval for the cycle counter corresponds to the sound best-case and worst-case execution time bounds, or BCET and WCET for short.

The example program in Fig.~\ref{fig:modexp_imp_cfg} contains two conditional jumps: line 6 and line 10.
The latter is for the loop iteration and is executed exactly 32 times, contributing a constant execution time.
The former depends on the bits of \registername{R0}, where taking the branch skips line 7.
When executing as in Sec.~\ref{nfm2023:app:symexec_app}, the two paths through the loop body have constant execution time and merge into an interval.
During composition and instantiation, the interval bounds are added together accordingly.
%
\todo{AL: drop side channel discussion to save space?}
Interestingly, equality of lower and upper bounds proves the absence of timing side channels.
While the example program is not free of side channels, we can make its execution time independent of \registername{R0}, e.g., balancing the execution times of the loop body branches by inserting nops.

\subsection{Evaluation}
\label{nfm2023:sec:eval}


We curated a set of evaluation programs, which we compiled with GCC unoptimized (except for the one marked with \verb|o3| and the assembly program \verb|ldldbr8|), tested on hardware, and then analyzed using both AbsInt aiT and HolBA-SE; the results are presented in Table~\ref{nfm2023:tbl:eval2}.
Concerning choice of programs, \texttt{ldldbr8} is a manually constructed counterexample to aiT soundness, \texttt{aes} showcases constant-time proofs via overapproximation to avoid state size explosion due to loops, \texttt{modexp} is the running example from Sec.~\ref{sec:background:example}, and \texttt{motor_set} and \texttt{pid} exemplify embedded applications where execution time may be an important factor.

Testing a program consists of generating 1000 random input states and collecting observed BCET and WCET when executing on an STM32F051R8T6 chip
measuring the clock cycles using the processor SysTick timer.
We use AbsInt aiT~\cite{AbsIntaiT} that is part of AbsInt ${\texttt{a}}^{\texttt{3}}$ for ARM Version 22.04.
It is based on abstract interpretation and is the industry standard for obtaining safe WCET bounds with static analysis, but does not provide BCET values.
For evaluations, aiT runs within 4 s for each program.
%
Testing and HolBA-SE count cycles as in the ARM specification of execution time for Cortex-M0~\cite{CortexM0instrcycles}, starting after the first instruction has been fetched and decoded.
The aiT tool instead assumes an initially empty pipeline, adding two or more clock cycles its analysis.

\begin{table}[t]
\begin{center}
\begin{tabular}{l|r|r||r|r||r||r|r|r}
\multicolumn{3}{c||}{\textbf{Program}} & \multicolumn{2}{c||}{\textbf{Testing}}  & \multicolumn{1}{c||}{\textbf{aiT}}    & \multicolumn{3}{c}{\textbf{\toolchain}}     \\ 
\textbf{name} & \textbf{\#instrs} & \textbf{\#splits} & \textbf{BCET} & \textbf{WCET}  & \textbf{WCET}    & \textbf{BCET} & \textbf{WCET} & \multicolumn{1}{c}{\textbf{eval}}     \\ \hline \hline 
ldldbr8          & 34 & 0            & 60 & 60       & 55$\ssymbol{2}$   & 60 & 60$\ssymbol{1}$ & 1 s      \\ \hline
aes              & 543 & 1           & 3761 & 3761   & 3761$\ssymbol{2}$ & 3761 & 3761$\ssymbol{1}$ & 8.4 min     \\ \hline
modexp           & 130 & 11          & 54358 & 93540 & 197721            & 43704 & 197560 & 11.4 min   \\
- uidivmod       & 85 & 9            & 1362 & 1778   & 3062              & 1332 & 3060 & 5.9 min          \\ \hline
motor_set        & 165 & 6           & 272 & 280     & 283               & 264 & 280 & 49 s           \\ \hline
motor_set o3     & 112 & 12          & 73 & 91       & 92$\ssymbol{2}$   & 68 & 96 & 89 s          \\ \hline
pid              & 2257 & 294        & 4589 & 5483   & 7521              & 1849 & 7529 & 61.8 min  \\
- fadd           & 401 & 70          & 77 & 155      & 188               & 68 & 178 & 12.4 min     \\
- fdiv           & 287 & 34          & 118 & 541     & 640              & 87 & 591 & 14.5 min       
\end{tabular}
\end{center}
\vspace{-0.5em}
\caption{\label{nfm2023:tbl:eval1}\label{nfm2023:tbl:eval2} Per program, the number of binary instructions, where splits are conditional or indirect jumps, BCET and WCET values are in clock cycles, and aiT only provides WCET. ${\@fnsymbol{2}}$ indicates unsound result and $\ssymbol{1}$ indicates constant time.
}
\vspace{-1.5em}
\end{table}






Program \verb|ldldbr8| consists of a sequence of 8 blocks, where each block is two loads and an unconditional branch to the next block, and a final nop.
As there is trivially only one path (no splits), execution time is constant.
%
The result illustrates that the pipeline in aiT can produce unsound results, and in contrast, the model in HOL4 is precise.
Further experiments have revealed that the pipeline modeled in aiT is not equivalent to the Cortex-M0 pipeline.
This has been confirmed with AbsInt along with the unsound result indicated with a ${\@fnsymbol{2}}$ in the table, which subsequently has been fixed in aiT.
Considering the different cycle counting of aiT, the programs \verb|aes| and \verb|motor_set o3| also represent counterexamples to the aiT analysis. 
However, HolBA-SE proves constant execution time for \verb|ldldbr8| and \verb|aes|, indicated using $\ssymbol{1}$.
%

Generally, the results of aiT and the symbolic execution are relatively similar.
The program \verb|motor_set| is as in Sec.~\ref{nfm2023:sec:impl:eval}, 
\verb|aes| performs AES encryption with 10 rounds, 
\verb|modexp| is the example from Sec.~\ref{sec:background:example}, and \verb|pid| is a larger case study for analyzing a compiled PID controller.
Since our processor does not support division and modulo, \verb|modexp| uses the inefficient but easily analyzable emulation function \verb|uidivmod|.
Similarly, \verb|pid| uses 11 floating-point operation emulation functions, provided as precompiled binaries with intricate control flows, like \verb|fadd| and \verb|fdiv|.






\section{Related Work}
\label{nfm2023:sec:relwork}

Symbolic execution~\cite{Baldoni2018,puasuareanu2009survey} was first introduced in the context of software testing by King~\cite{King1976} and applied to formal verification in follow-up work~\cite{Hantler1976}. Subsequently, it became widely used in both testing~\cite{David2016, Avgerinos2014, Saxena2009, Cha2012} and verification~\cite{Dannenberg1982, CoenPorisini2001}.
Concolic execution~\cite{Godefroid2005DART,stephens2016driller} underapproximates for performance and targets bug-finding, whereas we target verification and maintain soundness.

Our core rules (Sec.~\ref{nfm2023:app:symexec_app}) support standard techniques for making symbolic execution scalable. Array axiomatization~\cite{farinier2018arrays, Daniel2020, Cha2012} can flatten memory, enabling load–store bypassing and eliminating overwritten stores, curbing expression growth and solver load. Constant propagation reduces term size, while interval abstractions approximate expressions by bounded values~\cite{cousot1996abstract, Cha2012, hansen2009state}. State merging mitigates path explosion~\cite{Kuznetsov2012}, akin to widening in abstract interpretation~\cite{cousot1996abstract}. Interpolation reuses results by relaxing initial path conditions without introducing new paths~\cite{Jaffar2012Tracer} and value analysis can resolve indirect jumps~\cite{Kinder2009}.

In the context of binary verification, symbolic execution underpins translation validation~\cite{Currie2006, IosifLazar2015, Dasgupta2020}, validation of compiler optimizations~\cite{Tristan2008}, analysis of side-channel absence~\cite{Daniel2020, Daniel2021}, and test generation to validate side-channel models~\cite{Nemati2020, Buiras2021}, and data-flow/overflow analysis~\cite{Wang2009intscope}.
Many implementations rely on intermediate languages for architectural independence~\cite{bap, Wang2009intscope, Cha2012} but do not offer the theorem prover based trustworthiness of \toolchain.

Mazzucato~et~al.~\cite{Mazzucato2025} perform automated proofs in HOL Light of relational properties of ARMv8 and x86 machine code using Hoare logic. In their work, symbolic execution is one of many possible proof tactics; we provide a general symbolic execution theory independent of program logics and focus on proof automation for an intermediate language.
Sammler~et~al.~\cite{Sammler2022} import traces from SMT-based symbolic execution of binary code into Coq for reasoning; in contrast, our symbolic execution produces certified theorems inside HOL4.
Previous works typically use weakest preconditions to verify functional properties of unstructured code~\cite{Barthe2006,Lundberg2020,Barnett2005}, while our approach uses strongest postconditions.

WCET is fundamental in high assurance domains~\cite{Byhlin2005, Wartel2013, Souyris2007, Nowotsch2012, Kirner2005, Abella2015}. While we target a simple microarchitecture, prior work also abstracts complex timing features (e.g., caches, pipelines) to enable sound analysis~\cite{Heckmann2003, Li2006}.
%
Maroneze~et~al. extend the verified CompCert compiler with a verified loop-bound estimation~\cite{Maroneze2014}. However, they restrict to having a constant for WCET per instruction, which introduces overapproximation with our timing model.

\section{Conclusion}
\label{nfm2023:sec:conclusion}
We presented a theory of symbolic execution for unstructured programs suitable for theorem prover use. The theory builds on inference rules that are the basis of automated symbolic executors with common optimizations and adaptations for code using specific ISAs. We instantiated the theory for the BIR language, providing machine-checked proofs of soundness of symbolic execution of BIR in HOL4. Applications of our implementation~\cite{HolBA-SE} demonstrate proof automation in several case studies. Scaling further is possible via binary slicing and modular verification, but automation of such approaches is left to future work. 
\subsubsection*{Acknowledgments}
The authors thank Henrik Karlsson and Didrik Lundberg for their help with \toolchain case studies.
This work was partially supported by the Wallenberg AI, Autonomous Systems and Software Program (WASP) funded by the Knut and Alice Wallenberg Foundation and a gift from Intel.

%
%
%
\bibliographystyle{splncs04}
\bibliography{references}

\end{document}